\documentclass[12pt]{article}
\usepackage{graphics}
\begin{document}
\begin{center}
{The Self-energy of Nucleon for the Photon-proton Elastic Scattering \\ and the Electromagnetic Polarizability}
\end{center}
\begin{center}
{Susumu Kinpara}
\end{center}
\begin{center}
{\it Institute for Quantum Medical Science \\ Chiba 263-8555, Japan}
\end{center}
\begin{abstract}
The effect of the self-energy on the photon-proton elastic scattering is investigated for the backward and the forward directions.
The shape of the Thomson scattering at the photon energy $\omega \to 0$ is broken by taking into account the self-energy of proton.
The electromagnetic polarizabilities $\bar{\alpha}\pm\bar{\beta}$ are calculated 
by the lowest-order perturbative treatment. 
\end{abstract}
\section*{\normalsize{1 \quad Introduction}}
\hspace*{4.mm}
Nucleon is a fundamental element of the nuclear system and the motion is described by the nucleon propagator.
The exact form of the self-energy is required to understand the properties not only for the finite density
but also the zero density such as the scattering phenomena.
In free space with no medium effect the isolated nucleon mainly interacts with the virtual pions which are emitted and absorbed repeatedly. 
\\\hspace*{4.mm}
To comprehend the pion-nucleon system by means of the field theoretical method the interaction lagrangian is prepared
and the perturbative expansion is carried out.
The calculation of the higher-order corrections with the pseudovector coupling interaction appears to be impossible 
because the derivative coupling prevents us from removing the divergences 
in the framework of the counter terms for the mass and the wave function.
Then the non-perturbative approach is indispensable and yields the non-perturbative term added to the usual vertex part.
Consequently the algebraic equation is made feasible and it determines the self-energy keeping the usual way of the counter terms.
\\\hspace*{4.mm}
An interesting application is the electromagnetic property of the nucleon current.
According to the Ward-Takahashi (W-T) identity the vertex is connected with the form of the self-energy definitely.
The genuine structure and the validity of the approximation becomes clearer by comparing the results with the experimental data.
For example the lowest-order perturbative expansion has turned out that it overestimates the magnetic moment of nucleon.
The approximate form reproduces well assuming that the four-momentum of the incident photon is small.
It implies that the realistic structure is such that the diagrams of the higher-order perturbative corrections cancel 
the higher-order in the lowest approximation as the energy carried in decreases.
\\\hspace*{4.mm}
The higher energy region of the pion-nucleon scattering is useful to examine 
whether the energy dependence of the self-energy is valid or not.
It has been revealed that the non-perturbative term plays a decisive role to account for the dependence of the cross section 
on the incident energy and the charge of pion at the intermediate energy region \cite{Kinpara}.
Among the various models classified by the maximum order of the expansion 
the suitable one is obtainable by raising the order up to the third order.
The shift of the maximum order suggests us that the real form is dependent on the energy brought in the system
within the present study by the lowest-order approximation. 
The photon-proton elastic scattering is another phenomenon valuable for the investigation of the self-energy on the proton.
\section*{\normalsize{2 \quad The self-energy in the photon-proton elastic scattering}}
\hspace*{4.mm}
Many fundamental processes of proton are determined by the self-energy arising from the interaction with virtual pions.
Particularly the effect on the photon-proton elastic scattering is expected to describe both of the energy and the angular dependence.
The constant term of the amplitude expanded by the photon energy 
is independent of the dynamical properties according to the low energy theorem \cite{BD}.
Furthermore the static property also exists at the linear term which contains the constants of the charge, proton mass and the anomalous magnetic moment.
The appearance of the dynamical properties in the quadratic order indicates 
that the proton is not an isolated point-like object but has a composite structure of nucleon and pions.
The degree of the softness is expressed by the coefficients as the electric and the magnetic polarizabilities \cite{Petrun'kin}.
\\\hspace*{4.mm}
The perturbative expansion method is applied to the photon-proton elastic scattering analogous to the photon-electron case.
The lowest-order approximation is known as the Klein-Nishina formula for the compton scattering.
The correction is twofold.
By virtue of the W-T identity the photon-proton-proton vertex contains the self-energy 
and when both of two protons are on-shell state it generates the anomalous interaction added to the Dirac part.
The experimental value of the anomalous part of the magnetic moment of nucleon is reproduced by adjusting the pseudovector coupling constant 
in the lowest-order model.
The value of the anomalous coupling constant of proton ($\kappa \approx 1.79$) is used herein irrespective of the model.
\\\hspace*{4.mm}
To make another correction the intermediate state is extended.
The single proton state is constructed by including the state composed of one pion and nucleon.
Then the self-energy $\it\Sigma(q)$ is applied to the free propagator of nucleon. 
The explicit form has been obtained in our previous study.
It is given as a series of the power $a^n$ 
\begin{eqnarray}
\Sigma(q) = \frac{M a^2}{M^2+m_\pi^2-M a/2-a^2/2}= \sum_{n=2}^\infty \sigma_n(q) 
\end{eqnarray}
\begin{eqnarray}
a \equiv \gamma\cdot q - M
\end{eqnarray}
\begin{eqnarray}
\sigma_n(q) \equiv M \, c_1^{(n)}(q^2) - c_2^{(n)}(q^2) \, \gamma\cdot q
\end{eqnarray}
in terms of the coefficients $c_i^{(n)}(q^2)\;(i=1,2)$.
The $M$ is the proton mass and the quantity of the order $O(m_\pi^2/M^2)$ is neglected because the pion mass $m_\pi$ is much smaller than $M$.
The various models $\it\Sigma_n(q)$$\;(n=2,3,\cdots)$ are defined as the following 
\begin{eqnarray}
\Sigma_n(q) \equiv M \, \tilde{c}_1^{(n)}(q^2) - \tilde{c}_2^{(n)}(q^2) \, \gamma\cdot q
\end{eqnarray}
\begin{eqnarray}
\tilde{c}_i^{(n)}(q^2) \equiv \sum_{m=2}^n c_i^{(m)}(q^2)  \qquad  (i=1,2)
\end{eqnarray}
labeled by the number $n$.
\\\hspace*{4.mm}
Using the general form of the self-energy 
$\it\Sigma(q) = M c_{\rm 1}(q^{\rm 2})-c_{\rm 2}(q^{\rm 2})\gamma\cdot q$
the nucleon propagator $G(q) = (\,\gamma\cdot q \,-\,M \,-\,\Sigma(q)\,)^{-1}$ is expressed in the rationalized form as
\begin{eqnarray}
G(q) = a(q^2) \, \frac{\gamma\cdot q + M + \hat{M}(q^2)}{q^2 -M^2}
\end{eqnarray}
\begin{eqnarray}
a(q^2) \equiv (1+c_2(q^2))^{-1}\frac{q^2-M^2}{q^2-(M+\hat{M}(q^2))^2}
\end{eqnarray}
\begin{eqnarray}
\hat{M}(q^2) \equiv M \, \frac{c_1(q^2) - c_2(q^2)}{1 + c_2(q^2)}.
\end{eqnarray}
When the four-momentum $q$ is at the on-shell state $q^2=M^2$,
there exists the relation $c_1(M^2)=c_2(M^2)$ because of the renormalization of the proton mass.
The low energy limit of the cross section at the photon energy $\omega\rightarrow 0$ is calculated by the lowest-order approximation
along with the relation $dc_1(M^2)/dq^2-dc_2(M^2)/dq^2 = c_2(M^2)/2 M^2$ ascribed to the Ward identity \cite{Gell-Mann}.
Using the relations it is shown that the values of the differential cross section at the backward and the forward directions 
change from that of the Thomson scattering.
Accordingly the terms due to the dynamical properties should be cancelled 
by taking into account the other processes of the same order in the perturbative expansion.  
\\\hspace*{4.mm}
In order to examine the pole structure it is useful to express
the renormalized propagator $G(q)$ as $G(q)=(A(q^2)\gamma\cdot q +B(q^2) M)/(q^2-M^2)$ 
in which $A(q^2)\equiv a(q^2)$ and $B(q^2)\equiv a(q^2)(1+\hat{M}(q^2)/M)$.
At the on-shell state ($q^2=M^2$) the quantities $a(q^2)$ and $\hat{M}(q^2)$ are $a(M^2)=1$ and $\hat{M}(M^2)=0$ 
and the renormalization condition $A(M^2)=B(M^2)=1$ is satisfied.
The condition suffices to give the amplitude of the Thomson scattering at the low energy limit in the method of the low-energy theorem \cite{BD}. 
\\\hspace*{4.mm}
The propagator is applied to the evaluation of the differential cross section for the backward and the forward directions
of the scattering angle $\theta$ by the unpolarized proton target in the laboratory system.
It is given as
\\
\begin{eqnarray}
\frac{d \sigma}{d \Omega} 
=\frac{\alpha^2}{2^5 M^2}\,(\frac{\omega^\prime}{\omega})^2\,[\,\frac{T_1}{(k \cdot p)^2}
+\frac{T_2}{(k^\prime \cdot p)^2}+\frac{{\rm 2} \, T_3}{k\cdot p \, k^\prime \cdot p}\,]
\end{eqnarray}
\begin{eqnarray}
T_1 &\equiv& a((p+k)^2)^2 \,{\rm Tr}\,[(\gamma\cdot p^\prime+M) \gamma\cdot\epsilon^\prime \gamma\cdot\epsilon \,
\chi(k,k^\prime,\epsilon) \nonumber\\ && (\gamma\cdot p+M) \chi(k,k^\prime,\epsilon) 
\gamma\cdot\epsilon \gamma\cdot\epsilon^\prime]
\end{eqnarray}
\begin{eqnarray}
T_2 &\equiv& a((p-k^\prime)^2)^2 \,{\rm Tr}\,[(\gamma\cdot p^\prime+M) \gamma\cdot\epsilon \gamma\cdot\epsilon^\prime
\chi(-k^\prime,-k,\epsilon^\prime) \nonumber\\ && (\gamma\cdot p+M) \chi(-k^\prime,-k,\epsilon^\prime)
\gamma\cdot\epsilon^\prime \gamma\cdot\epsilon]
\end{eqnarray}
\begin{eqnarray}
T_3 &\equiv& -a((p+k)^2)a((p-k^\prime)^2) \,{\rm Tr}\,[(\gamma\cdot p^\prime+M) \gamma\cdot\epsilon^\prime
\gamma\cdot\epsilon \chi(k,k^\prime,\epsilon) \nonumber\\ 
&& (\gamma\cdot p+M) \chi(-k^\prime,-k,\epsilon^\prime)\gamma\cdot\epsilon^\prime \gamma\cdot\epsilon]
\end{eqnarray}
\begin{eqnarray}
&&\chi(k,k^\prime,\epsilon) \equiv (1-2b\gamma\cdot k^\prime-4b k^\prime\cdot\epsilon\gamma\cdot\epsilon)\nonumber\\
&&\times\left[\,\{1+2 b(2 M + \hat{M}((p+k)^2))\}\gamma\cdot k-\hat{M}((p+k)^2)-4 b p\cdot k \,\right]
\end{eqnarray}
\\
where $\alpha$ is the fine structure constant.
The $\omega$, $\omega^\prime$ in the kinematic factor $(\omega^\prime/\omega)^2$ are the absorbed and the emitted photon energies.
These variables are connected by the relation
\begin{eqnarray}
\omega^\prime = \frac{\omega}{1+\omega(1-{\rm cos} \theta)/M}
\end{eqnarray}
under the conservation of the four-momentum $k+p = k^\prime + p^\prime$
where $k$, $p$, $k^\prime$ and $p^\prime$ are the incident photon, the target proton, the scattered photon and the recoil proton respectively.  
By virtue of the self-energy term in the W-T identity the vertex includes the anomalous interaction.
It generates the anomalous part of the magnetic moment with $b \equiv \kappa/4 M$ and $\kappa$ = 1.79.
Then the simple form of $\chi(k,k^\prime,\epsilon) = \gamma\cdot k -\hat{M}$ is modified by the tensor component $\sigma_{\mu \nu}$ of the gamma matrix.
\\\hspace*{4.mm}
When the direction of $k^\prime$ is either the backward ($\theta = \pi$) or the forward ($\theta = 0$) 
the number of the $\gamma$ matrix in the trace is at most eight 
and the $\chi(k,k^\prime,\epsilon)$ remains the linear form in $\gamma\cdot k$.
It is expressed by the quantity $\chi(k,\omega,\omega^\prime)$ identical to $\chi(k,k^\prime,\epsilon)$ as
\begin{eqnarray}
\chi(k,\omega,\omega^\prime) = F(\omega,\omega^\prime)\,(\gamma\cdot k -\tilde{M}(\omega,\omega^\prime))
\end{eqnarray}
\begin{eqnarray}
&&F(\omega,\omega^\prime) \equiv 1+4 M b \nonumber\\
&&-2 b \,[ 4Mb\,\omega^\prime +(\omega^\prime \omega^{-1}-1)\hat{M}(\omega)
-2 \omega^\prime(1+4 M b + 2 b \hat{M}(\omega))\,]
\end{eqnarray}
\begin{eqnarray}
\tilde{M}(\omega,\omega^\prime) \equiv F(\omega,\omega^\prime)^{-1} \,
[\,(1-4 \,b \,\omega^\prime)(\hat{M}(\omega)+4 M b \,\omega) \nonumber\\
+\,8 \,b \, \omega \omega^\prime (1+4 M b +2 b \hat{M}(\omega))\,]
\end{eqnarray}
for the $\theta = \pi$ scattering and
\begin{eqnarray}
F(\omega,\omega^\prime) \equiv 1+4 b (M+\hat{M}(\omega)+2 M b \,\omega)
\end{eqnarray}
\begin{eqnarray}
\tilde{M}(\omega,\omega^\prime) \equiv  F(\omega,\omega^\prime)^{-1} (\hat{M}(\omega)+4 M b \,\omega)
\end{eqnarray}
for the $\theta = 0$ scattering.
\\\hspace*{4.mm}
Taking the summation over the polarization vector of the final photon ($\epsilon^\prime$) and averaging the initial photon ($\epsilon$) 
the differential cross section is as follows
\begin{eqnarray}
\frac{1}{2}\sum_{\epsilon\,\epsilon^\prime}\frac{d \sigma}{d \Omega}
= \frac{\alpha^2}{2 M^2}(\frac{\omega^\prime}{\omega})^2 \, J
\end{eqnarray}
\begin{eqnarray}
&&J \equiv a(\omega)^2 F(\omega,\omega^\prime)^2 \,[ \, \frac{\omega^\prime}{\omega} 
-2(1-\frac{\omega^\prime}{M})\frac{\tilde{M}(\omega,\omega^\prime)}{\omega}
+(1+\frac{\omega-\omega^\prime}{2 M})\frac{\tilde{M}(\omega,\omega^\prime)^2}{\omega^2} \,]\nonumber\\
&&+\;(\omega\leftrightarrow -\omega^\prime).
\end{eqnarray}
When the self-energy of the propagator is set equal to zero ($a(\omega)\equiv 1$ and $\hat{M}(\omega)\equiv 0$),
it reduces to that of the point-like proton having the anomalous interaction.
In fact the form of the vertex is not exact when either of the nucleons is at the off-shell state. 
\section*{\normalsize{3 \quad The effect of the self-energy on the polarizabilities}}
\hspace*{4.mm}
Besides the kinematic factor $(\sim (\omega^\prime/\omega)^2)$ in Eq. (20) the differential cross section is  
expressed by the power series of $x\equiv\omega\omega^\prime/M^2$ as shown in the Appendix 1.
At $\omega, \,\omega^\prime \rightarrow 0$ only the lowest-order term of the expansion remains 
as $J \to J_0 \equiv 2(1+c_1(M^2))^{-2}$ contrary to the low-energy theorem.
While the low-energy limit of the compton scattering in the present formulation 
is independent of the parameter $\kappa$ like the Thomson scattering the dynamical effect of the self-energy does not vanish
and changes the value of the limit.
Then the joint effect of the real vertex and the photon-photon-proton-proton 4-point vertex is required to satisfy the low-energy theorem. 
\\\hspace*{4.mm}
At present our interest of $J$ is the linear term of the power series expansion $J=J_0 + \lambda x + O(x^2)$.
The self-energy of the internal propagator changes the value of the coefficient from $\lambda$ to $\lambda+\lambda^\prime$.
The displacement $\lambda^\prime$ from that of the point-like proton case is given as
\begin{eqnarray}
\lambda^\prime = - \frac{M^3}{\alpha} 
[\, (\bar{\alpha}+\bar{\beta})(1+{\rm cos}\,\theta)^2+(\bar{\alpha}-\bar{\beta})(1-{\rm cos}\,\theta)^2 \,]
\end{eqnarray}
in terms of the electric and magnetic polarizabilities $\bar{\alpha}$ and $\bar{\beta}$ \cite{Olmos}.
The non-zero values mean that the proton certainly has an internal structure and they are indices of the softness.
In the present calculation it is interpretted as the mixing of the intermediate state composed of one boson and one fermion.
\\\hspace*{4.mm}
The $\bar{\alpha} \pm \bar{\beta}$ are given by the relation as
\begin{eqnarray}
\bar{\alpha} \pm \bar{\beta} 
= -\frac{\alpha}{4 M^3} \lim_{x \to 0} x^{-1}[\,J_{\pm}-J_0 -x\{ (1\mp1)^2 +a_0 \pm a_1 +a_2 \}\,]
\end{eqnarray}
where the values of $a_i$ ($i\,$= 0,1,2) are seen in Ref. \cite{Olmos}.
The subscript $+/-$ of $J_{\pm}$ denotes the $J$ for the $\theta = 0/\pi$ directions in Eq. (21). 
\\\hspace*{4.mm}
The dependence of $J_\pm$ on $x$ is determined by $a(\omega)$ and $\hat{M}(\omega)$ which constitutes the self-energy.
The form is given in the Appendix 2.
As we have seen in our study of the electromagnetic form factor of nucleon and the pion-proton scattering
the exact form does not necessarily provide the optimum values for these energies.
Then the various models ($n\,=\,2,\cdots, \infty$) are compared with each other to infer the best form of $\it\Sigma(k)$.
\\\hspace*{4.mm}
The backward scattering plays an important role since the process is explained by the perturbative expansion
and it is sensitive to the assumed shape of the intermediate state.
In Table 1 difference between the electric and magnetic polarizability $\bar{\alpha} - \bar{\beta}$
is shown for the various models of $n=2,3,4,5,6,\infty$ in units of 10${}^{-4}\,{\rm fm}^{3}$.
They are nearly the same magnitude and close to the experimental value 
$\bar{\alpha} - \bar{\beta}=7.03 \,\times$10${}^{-4}\,{\rm fm}^{3}$ \cite{Zieger}.
\begin{center}
\begin{tabular}{|c|c c c c c c|}
   \multicolumn{7}{c}{ Table 1 }\\
      \hline
 n & 2 & 3 & 4 & 5 & 6 & $\infty$ \\
      \hline
 $\bar{\alpha} - \bar{\beta}$ & 6.46 & 6.68 & 7.15 & 9.63 & 7.64 & 5.98 \\
     \hline
 $\bar{\alpha} + \bar{\beta}$ & -2.67 & -3.18 & -3.20 & -13.2 & -3.69 & -2.47 \\
     \hline
\end{tabular}
\end{center}
\hspace*{4.mm}
The behavior of the cross section at the low energy limit of the photon energy $\omega\to 0$ enables us to select the suitable one.
Among the various models the $n$=3 is distinct from the others.
The low energy limit does not depend on the dynamical property and returns to the original form since $c_1(M^2)$=0 neglecting the pion mass.
It is noted that the pion mass shifts the value of $\bar{\alpha} - \bar{\beta}$ from 6.68 to 6.63 only about one percent.
\\\hspace*{4.mm}
For the forward scattering the procedure of the dispersion theory is effective and it gives the main contribution to $\bar{\alpha} + \bar{\beta}$.
The value of it is $\bar{\alpha} + \bar{\beta}=13.8\,\times$10${}^{-4}\,{\rm fm}^{3}$ \cite{Olmos}.
Meanwhile all of the perturbative calculations with the inclusion of the self-energy encounters the unsatisfactory results.
It may be ascribed to replacement of the photon-proton-proton vertex
with the interaction of the anomalous magnetic moment and lack of two-photon vertex arising from the self-energy
as well as the lowest-order term ($\sim\omega^{0}$).
There is no remarkable difference in the numerical value except for the $n$=5 model and it resembles to the tendency for the backward scattering.
\\\hspace*{4.mm}
The low-energy theorem is useful to calculate the forward scattering amplitude.
It makes us possible to include the two-photon vertex by differentiating the self-energy without considering the diagram in detail.
All of $\bar{\alpha} + \bar{\beta}$ results in an order of the magnitude smaller than the experimental value with the correct sign.
In order to obtain the appropriate value of $\bar{\alpha} + \bar{\beta}$ 
it is reasonable to examine the scattering of photon by the pion propagator in the self-energy.
The seagull diagram that is the process through the photon-photon-pion-pion 4-point vertex is selected 
and which may be calculated by the method of the low-energy theorem along with the substitution
\begin{eqnarray}
\frac{\partial^2}{\partial p_\mu \partial p_\nu} \Sigma(p+k) \to -2 g^{\mu\nu} \frac{\partial}{\partial m_\pi^2} \Sigma(p+k)
\end{eqnarray}
where $p$ and $k$ are the four-momenta of the proton and the photon.
The power series expansion in the photon energy $\omega$ is quadratic at $\omega \to 0$. 
The coefficients of the models have the divergent tendency as the label number increases and the respective signs are alternate.
Although the approximate way in Eq. (24) neglects the usual process by two successive pion-photon-pion vertices 
it attracts our interest of the $\sim \omega^2$ term in the forward direction.
Among the models the $n$=5 case shows $\bar{\alpha} + \bar{\beta}=13.4\times$10${}^{-4} \,{\rm fm}^3$
in agreement with the experimental value.
There may be a connection between the advantage of it staying at the experimental area and appearance of the peak in Table 1.
\section*{\normalsize{4 \quad Summary }}
\hspace*{4.mm}
The photon-proton elastic scattering has been studied by taking into account the self-energy of the propagator.
Because the form has still not been determined within the present our stage 
the various models classified by the order of the expansion are suggested to evaluate the polarizabilities of proton.
One of the methods to choose the suitable model is to compare the low energy limit.
Among the models the $n$=3 one almost does not move from the value of the Thomson scattering cross section.
The usefulness of it is similar to the pion-proton scattering 
in which the ratio of the total cross section between $\pi^{+}$ and $\pi^{-}$ is described well by the model.
For the forward direction the self-energy of proton works destructively.
Therefore it indicates that the interaction of photon with the virtual pions around the proton is essential
and achieves the description of the dynamical properties by the $n$=5 model.   
\\\hspace*{4.mm}
\section*{\normalsize{ Appendix 1}}
\hspace{4.mm}
At $\omega$, $\omega^\prime$ $\to$ 0 the $J$ in Eq. (21) is expanded in powers of $\omega$ and $\omega^\prime$.
The respective term is expressed by
\begin{eqnarray}
{\omega^\prime}^n \omega^{-m} + (\omega \leftrightarrow -\omega^\prime) 
=(\omega\omega^\prime)^{-m}({\omega^\prime}^{m+n}+(-1)^{m+n}\omega^{m+n})\nonumber
\end{eqnarray}
where the $n$ and $m$ are the integer values.
Here the $\omega$ and $\omega^\prime$ divided by the proton mass
are used to make the equations dimensionless.
It is sufficient to consider the part
\begin{eqnarray}
{\omega^\prime}^{N}+(-1)^{N}\omega^{N} \qquad(N \equiv m+n)\nonumber
\end{eqnarray}
for the factor in front of it is $x^{-m}$.
When $N < {\rm 0}$ it is converted to 
\begin{eqnarray}
(-1)^{N} x^{N} ({\omega^\prime}^{-N}+(-1)^{N} \omega^{-N})\nonumber
\end{eqnarray}
and then $N$ is set to $N > 0$ from the outset.
By expanding $(a+b)^N$ the general relation is obtained as
\begin{eqnarray}
(a+b)^N = a^N+b^N +\sum_{k=1}^{[\frac{N-1}{2}]}{}_NC_k(ab)^k(a^{N-2k}+b^{N-2k})
+\mit\Delta\,{}_NC_{\frac{N}{\rm 2}}(ab)^{\frac{N}{\rm 2}}\nonumber
\end{eqnarray}
\begin{eqnarray}
\mit\Delta \equiv \frac{\rm 1+(-1)^{\it N}}{\rm 2}.\nonumber
\end{eqnarray}
Substituting $a=\omega^\prime$ and $b=-\omega$ the recursion relation yields
\begin{eqnarray}
{\omega^\prime}^{N}+(-1)^{N}\omega^{N} 
= ({\rm cos}\theta-1)^N x^N-\mit\Delta\,{}_NC_{\frac{N}{\rm 2}}(-x)^\frac{N}{\rm 2}\nonumber\\
-\sum_{k=1}^{[\frac{N-1}{2}]}{}_NC_k(-x)^k({\omega^\prime}^{N-2k}+(-1)^{N-2k}\omega^{N-2k})\nonumber
\end{eqnarray}
using the relation $\omega^\prime-\omega = ({\rm cos}\theta-1)x$.
\section*{\normalsize{ Appendix 2}}
\hspace{4.mm}
Using the sum of the initial four-momentum of proton ($p$) and photon ($k$) $q \equiv p+k$ 
the argument of $a(q^2)$ and $\hat{M}(q^2)$ is given as $q^2 = M^2 +2 M \omega$ where $\omega$ is the laboratory energy of the incident photon.
The following quantities $a(\omega)$ and $\hat{M}(\omega)$ are defined as $a(\omega) \equiv a(q^2)$ and $\hat{M}(\omega) \equiv \hat{M}(q^2)$.
For $n$=2, 3, 4, 5, 6 and $\infty$ the $a(\omega)$ and $\hat{M}(\omega)$ are presented.
The pion mass is dropped approximately ($m_\pi$=0).
\\\\
$n$=2)
$\qquad 
a(\omega)=(1-\frac{2 \eta}{3})^{-1}
\qquad 
\hat{M}(\omega)=2 M \eta/3$
$\qquad(\,\eta \equiv \omega/M \,)$
\\\\
$n$=3)
$\qquad 
a(\omega)=(1-\eta)^{-1}
\qquad 
\hat{M}(\omega)=0$
\\\\
$n$=4)
$\qquad 
a(\omega)=\frac{2 (7+5 \eta)}{14+2 \eta-16 \eta^2-9 \eta^3}
\qquad 
\hat{M}(\omega)=M \eta \frac{6+3 \eta}{7+5 \eta}$
\\\\
$n$=5)
$\qquad 
a(\omega)=\frac{2(6+20 \eta+5 \eta^2)}{12+36 \eta-28 \eta^2-32 \eta^3-25 \eta^4}
\qquad 
\hat{M}(\omega)=M \eta \frac{4+7 \eta}{3+10 \eta+\frac{5}{2} \eta^2}$
\\\\
$n$=6)
$\qquad 
a(\omega)=\frac{8(19+34 \eta+14 \eta^2)}{152+192 \eta-164 \eta^2-288 \eta^3-192 \eta^4-121 \eta^5}
\qquad 
\hat{M}(\omega)=M \eta \frac{18+26 \eta+\frac{11}{2}\eta^2}{19+34 \eta+14 \eta^2}$
\\\\
$n$=$\infty$)
$\qquad 
a(\omega)=\frac{4-5 \eta+2 \eta^2}{4-7 \eta+2 \eta^2}
\qquad 
\hat{M}(\omega)=M \eta\frac{4(1-\eta)}{4-5\eta+2\eta^2}$
\\
\end{document}